\def\year{2021}\relax
\documentclass[letterpaper]{article} 
\usepackage{aaai21}  
\usepackage{times}  
\usepackage{helvet} 
\usepackage{courier}  
\usepackage[hyphens]{url}  
\usepackage{graphicx} 
\usepackage{natbib}  
\usepackage{caption} 
\usepackage{hyperref}
\title{Improving Perceptual Quality of Drum Transcription \\
      with the Expanded Groove MIDI Dataset}
\author {
        Lee Callender,\thanks{Equal contribution}
        Curtis Hawthorne,\footnotemark[1]
        Jesse Engel \\
}
\affiliations {
    Google \\
    leefcallender@gmail.com, fjord@google.com, jesseengel@google.com
}
\nocopyright

\begin{document}

\maketitle

\begin{abstract}
We introduce the Expanded Groove MIDI dataset (E-GMD), an automatic drum transcription (ADT) dataset that contains 444 hours of audio from 43 drum kits, making it an order of magnitude larger than similar datasets, and the first with human-performed velocity annotations. We use E-GMD to optimize classifiers for use in downstream generation by predicting expressive dynamics (velocity) and show with listening tests that they produce outputs with improved perceptual quality, despite similar results on classification metrics. Via the listening tests, we argue that standard classifier metrics, such as accuracy and F-measure score, are insufficient proxies of performance in downstream tasks because they do not fully align with the perceptual quality of generated outputs.
\end{abstract}

\section{Introduction}
Discriminative models predict the conditional distribution $p(y|x)$ over labels $y$ that correspond to an input $x$.  In the space of automatic drum transcription (ADT), discriminative models are used to predict when and what drum hits are used in a drum performance conditional on audio input of a performance. 

While classifier metrics such as accuracy, precision, recall, and F-measure scores are often used to evaluate discriminative models, decision theory highlights that the true quantity of interest is the expected utility (or cost) of the inferred labels in a downstream task~\citep{von2007theory}.



Recent work on piano transcription has demonstrated the value of considering downstream generation, showing that separately classifying note onsets from note persistence led to dramatic improvements in the perceptual quality of generation due to a reduction in false positive onsets~\citep{hawthorne2018onsets}. For the application of drum transcription, we develop a new dataset and transcription model capable of transcribing drum hit velocity (loudness) and examine how that capability contributes to the perceived quality of the transcriptions.


Our key contributions include:

\begin{itemize}
\itemsep0em
\item The Expanded Groove MIDI dataset (E-GMD), the first dataset to capture both expressive timing and velocity of human performances and a dataset size that is an order of magnitude larger than similar datasets.
\item Training expressive ADT models on E-GMD to predict timings, drum hit, and velocity by incorporating a separate velocity-prediction head.
\item Demonstrating that predicting expressive dynamics (velocity) in addition to timing generates outputs with improved perceptual quality, as determined by listening tests, despite achieving similar results on classification metrics.
\item Developing a new \textit{Shuffled mixup} strategy for data augmentation and regularization that effectively limits overfitting.
\end{itemize}

Audio samples of the dataset and examples used in the listening test are provided in the online supplement at \url{https://goo.gl/magenta/e-gmd-examples}, and the full dataset is available at \url{https://g.co/magenta/e-gmd} under the Creative Commons Attribution 4.0 International (CC BY 4.0) license.

\section{Related Work}
The recent work of Wu et al.~\shortcite{review} provides a comprehensive overview of ADT and includes evaluation of current state of the art methods. While there has been a large collection of studies published over ADT in recent years~\cite{multi, choi2019deep, Cartwright2018INCREASINGDT, chih_wei_wu_2018_1492447, Carl_Southall_2018, southall2018player, ueda2019bayesian}, most ADT research has maintained a focus on classifier metrics to assess quality.

Of the approaches that have explored deep learning~\cite{multi, choi2019deep, Cartwright2018INCREASINGDT, Carl_Southall_2018}, research is still fairly new given the large data required to effectively produce a model. As annotating drums is still a fairly manual task, most datasets for ADT are relatively small in size and resource intensive to create.  This has lead to new research into solving that problem, including unsupervised approaches~\cite{choi2019deep, chih_wei_wu_2018_1492447} and the creation of synthetic datasets~\cite{choi2019deep, multi, Cartwright2018INCREASINGDT, Miron2013AnOD}.

Given the difficulty of ADT and the limited datasets available, the overwhelming majority of ADT research has focused on ADT with the classification of 3 primary drum hits: Kick Drum, Snare Drum, Hi-hat (KD, SN, HH)~\citep{dittmar2014real, lindsay2012drumkit, wu2015drum, vogl2016recurrent, vogl2017drum, stables2016automatic, Southall2017AutomaticDT}. A handful of datasets contain annotations beyond the 3 standard hits, however the set of drum hits is not standardized, with each dataset containing a varied collection of drum hits~\citep{multi, Cartwright2018INCREASINGDT, Dittmar04furthersteps}.

Velocity has sometimes been considered during ADT tasks. For example, in DrummerNet~\cite{choi2019deep}, velocity is used as a probability of hit for peak-picking. However, velocity is not predicted as part of overall model output. To the best of our knowledge, our work is the first model that directly predicts velocity values and evaluates the perceptual quality of resynthesized outputs.

\begin{table}[ht]
\centering
\begin{tabular}{lcccc}
\hline
Dataset & Minutes & Kits & Human & Vel \\
\hline
E-GMD & 26,670 & 43 &  $\surd$ & $\surd$ \\
TMIDT & 15,540 & 57 & $\times$ & $\times$ \\
IDMT & 130 & 6 & $\times$ & $\times$ \\
ENST & 61 & 3 & $\surd$ & $\times$ \\
MDB Drums & 21 & $\approx$23 & $\surd$ & $\times$ \\
RBMA13 & 103 & $\approx$30 & $\surd$ & $\times$ \\
\hline
\end{tabular}
\caption{Comparison of public datasets for ADT, including whether they contain exclusively human performances and velocity annotations. The exact number of kits in MDB Drums and RBMA13 is unclear, but is unlikely to exceed the total number of tracks, which is 23 and 30 respectively. All datasets contain isolated drum tracks, with the exception of RBMA13.}
\label{datasets}
\end{table}

\section{Datasets}
Only a handful of public datasets are available for ADT, and many have limited size and diversity. An even smaller subset of datasets contain human performances, and no public datasets contain human performances with velocity annotations~\cite{Cartwright2018INCREASINGDT,review,multi}. Reasons for these limitations include the tedious nature of generating labels for real drum performances and restrictions around licensing and intellectual property.

The difficulty of annotating real drum performances has inspired some recent studies to generate their own synthetic datasets. These datasets are commonly generated by taking a collection of MIDI (Music Instrument Digital Interface, the industry standard format for symbolic music data) drum performances and synthesizing audio via drum samples~\cite{Miron2013, multi, Cartwright2018INCREASINGDT}. Only one of these datasets is public~\cite{multi}, and it does not contain velocity annotations.

Table~\ref{datasets} compares several public datasets, including E-GMD. Of these datasets, we decided to use IDMT-SMT~\cite{Dittmar2014RealTimeTA} and ENST~\cite{enst} in our evaluations because of their commonality in prior studies. We opted not to use MDB Drums~\cite{southall2017mdb} because of its small size and did not use the dataset from Vogl et al.~\shortcite{multi}, which we refer to as TMIDT, because the licensing of its source material was ambiguous. We also did not use RBMA13~\cite{Vogl2017DrumTV} because the tracks included music in addition to drums, and we focused on transcribing only solo drumming.

E-GMD has many different annotated hits. For evaluation and listening tests, we group the annotated hits down to a 7 and 3 hit classification task, as shown in Table~\ref{drum-hierarchy-table}.

\begin{table}[ht]
\centering
\begin{tabular}{|l|c|c|}
\hline
E-GMD Hits & 7 hit & 3 hit\\
\hline
Kick drum & KD & KD \\
\hline
Snare drum &  & \\
Snare rim & SD & \\
Cross-stick & & \\
Clap & & \\
\cline{1-2}
Tom 1 &  & SD\\
Tom 1 Rim & &\\
Tom 2 & TT & \\
Tom 2 Rim & & \\
Tom 3 & &\\
Tom 3 Rim & &\\
\hline
Open Hi-Hat & & \\
Open Hi-Hat Bow & &\\
Closed Hi-Hat Bow & HH & \\
Closed Hi-Hat Bow & & \\
Pedal Hi-Hat & & \\
Tambourine & & \\
\cline{1-2}
Crash 1 Bow & & HH \\
Crash 1 Edge & CY & \\
Crash 2 Bow & & \\
Crash 2 Edge & & \\
\cline{1-2}
Ride Bow & RD & \\
Ride Edge & & \\ 
\cline{1-2}
Ride Bell & BE & \\
Cow Bell & & \\
\hline
\end{tabular}
\caption{The drum hit hierarchy for E-GMD. The 3 and 7 hit groupings are used in our model for evaluation and the listening test.}
\label{drum-hierarchy-table}
\end{table}

\subsection{IDMT-SMT}
IDMT-SMT contains only the 3 standard drum hits (KD, SN, HH), and contains 4 different drum kits. The dataset uses relatively simple drum patterns and contains audio and ground truth hit annotations. One drum kit is an acoustic kit that was recorded with varying velocities, however the ground truth annotations do not consider velocity and only consider drum hit type and timing. The other 3 drum kits use synthesized drums. The dataset contains audio for both individual hits and the mix of 3 hits. We use the full audio mix recordings for evaluation, and use the entire dataset because it is limited in length.

\subsection{ENST}
The ENST dataset was recorded with three different acoustic drum kits, performed by three professional drummers. Each performer used either sticks, rods, brushes, or mallets for each sequence, to produce a variety of timbres.

The dataset contains audio of single instrument strokes, short phrases, and drum tracks with and without additional accompaniment. The annotations contain labels for 20 different drum hits. While the performances for ENST are recorded, there again is no velocity annotation.

For our experiments, the tracks of isolated drum performances were used (the tracks labeled ``minus-one"), which is consistent with the other ADT studies we compare against. These isolated drum performances make up 64 tracks of 61s average duration and a total duration of 1 hour. We use all 64 tracks in evaluation. The rest of the dataset (single strokes, patterns) is ignored.

\subsection{Expanded Groove MIDI Dataset}
We introduce an expansion of the Groove MIDI Dataset (GMD), which we call the Expanded Groove MIDI Dataset (E-GMD). GMD is a dataset of human drum performances recorded in MIDI format on a Roland TD-11\footnote{\url{https://www.roland.com/us/products/td-11/}} electronic drum kit, and was originally created for generative drum sequencing~\cite{2019arXiv190506118G}.
MIDI information includes events like notes, that associate instrument, a time and a velocity together as an event.

GMD contains 13.6 hours, 1,150 MIDI files, and 22 different drum instruments. The dataset additionally includes synthesized audio outputs of the TD-11 aligned within 2ms of the corresponding MIDI files. The data includes performances by a total of 10 drummers, 5 professionals and 5 amateurs, with more than 80 percent coming from the professionals. The professionals were able to improvise in a wide range of styles, resulting in a diverse range of performances.

To make the dataset applicable to ADT, we expanded it by recording 43 drumkits on a Roland TD-17\footnote{The TD-17 is an award-winning electronic drum kit that ``faithfully reproduces the character and tone of acoustic drums." \url{https://www.roland.com/us/products/td-17_series/}}, ranging from electronic (e.g., 808, 909) to acoustic sounds. The additional drumkits were recorded at 44.1kHz and 24 bits and aligned within 2ms of the original MIDI files.
Using the Roland TD-17, a close analog to the Roland TD-11 (no longer manufactured) used in the original Groove dataset, enables accurate reproduction of nuances in the initial performances.  

We implemented a semi-manual process to systematically record new audio from the TD-17. The audio was recorded in real-time on a Digital Audio Workstation (DAW) and took about 16 hours to complete per kit. Given the semi-manual nature of the pipeline, there were some errors in the recording process that resulted in unusable tracks. The final numbers for E-GMD are shown in Table~\ref{egmd-stats}.

We maintained the same train, test and validation splits across sequences that GMD had. As each kit was recorded for every sequence, we see all 43 kits in the train, test and validation splits. The count of hits across all splits is shown in Table~\ref{egmd-hit-counts}.     

\begin{table}[ht]
\centering
\begin{tabular}{lcccc}
\hline
Split & Unique Seq & Total Seq & Dur \\
\hline
Train & 819 & 35,217 & 341.4h \\
Test & 123  & 5,289 & 50.9h \\
Validation & 117 & 5,031 & 52.2h \\
\hline
Total & 1,059 & 45,537 & 444.5h \\
\hline
\end{tabular}
\caption{E-GMD unique sequences, total sequences, and duration in hours by split.}
\label{egmd-stats}
\end{table}

\begin{table}[ht]
\centering
\begin{tabular}{lcccc}
\hline
Hit & Train & Test & Validation\\
\hline
KD & 2,181k & 319k & 343k \\
SD & 3,477k  & 468k & 533k \\
HH & 3,045k & 553k & 518k \\
TT & 805k & 98k & 171k \\
RD & 1,260k & 105k & 84k\\
BE & 191k & 9k & 21k \\
CY & 122k & 10k & 27k \\
\hline
\end{tabular}
\caption{E-GMD hit counts across splits in thousands. We show the counts for the seven hit grouping of E-GMD for brevity. See Table~\ref{drum-hierarchy-table} for hit definitions and grouping description.}
\label{egmd-hit-counts}
\end{table}

The online supplement includes examples of different sequences and kits at \url{https://goo.gl/magenta/e-gmd-examples}. The dataset is available at \url{https://g.co/magenta/e-gmd} under the Creative Commons Attribution 4.0 International (CC BY 4.0) license. The model described in this paper was trained with the v1.0.0 release of the dataset.

\section{Model}
We base our model on Onsets and Frames~\cite{hawthorne2018onsets} and adapt its note and velocity prediction capabilities to drum hit and velocity predictions. We call our new model OaF-Drums. 

\begin{figure}[ht]
\centering
\includegraphics[width=.8\columnwidth]{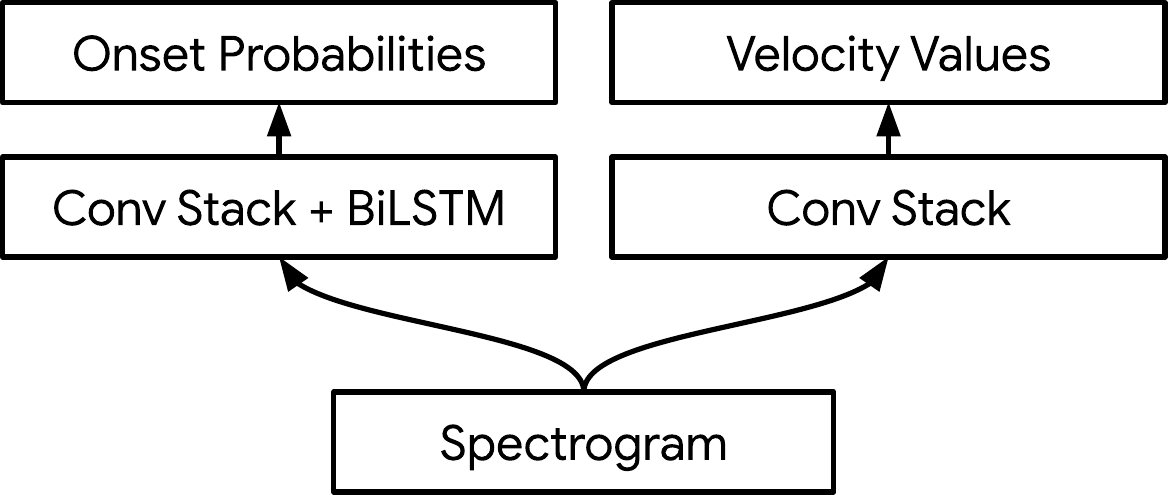}
\caption{OaF-Drums Model Architecture}
\label{architecture}
\end{figure}

We use only the onset and velocity stacks of the network, as illustrated in Figure~\ref{architecture}, because drum hits do not sustain like piano notes and so we do not require the frame or offset predictions. Complete network details are given in the Supplement.

For log mel-spectrogram creation, we increased the audio sample rate from 16 KHz to 44.1 KHz, the number of bins from 229 to 250, and shortened the hop length from 512 to 441 samples, resulting in frames with a 10ms width~\cite{multi}. We found the higher sample rate improved the model's ability to process events with high-frequency content like cymbal crashes, and the higher frame resolution was important for predicting events that repeated rapidly, such as drum rolls. The resulting higher resolution network required more memory during training, so we also switched from processing batches of 20-second segments to 12-second segments.

For labels, we forced onset labels to occupy a single frame instead of being spread across 30ms of frames as they are in the original piano model. This also helped improve accuracy for rapidly repeating events. Finally, we added a 0.5 weight multiplier to the velocity loss to prioritize correct hit recognition during training.

We found that overfitting on the training data was a significant concern. The initial manifestation of this problem was that the trained model would transcribe only the first and last few seconds of an evaluation sequence. We suspect this was due to the bidirectional LSTM layer memorizing drum sequences that are simpler than the piano sequences this architecture was originally designed for (8 hits instead of 88 notes). Also, even though our training data has 35,217 audio examples due to our many drum kits, there are only 1,059 unique drum hit sequences.

To prevent overfitting, we used the standard techniques of reducing model capacity and adding dropout~\citep{merity2017regularizing}. We decreased the size of the bidirectional LSTM layer from 128 to 64 units and added dropout at a rate of 50\% to the outputs of the LSTM cells, but this alone was insufficient.

We also used a form of \textit{mixup}~\cite{zhang2017mixup} for data augmentation and regularization. We created 500,000 training examples by randomly selecting pairs of examples from the training set, repeating the shorter of the examples until it was as long as the longer one, and then mixing their audio samples and underlying MIDI data together (prior to spectrogram or piano roll calculation) to form a new example, which is then split into 12-second chunks. This improved evaluation scores, but we still saw strongly divergent train/evaluation curves.

To create further diversity during training, we split those 500,000 examples into 1-second chunks. Then, at training time we spliced together random chunks into a 12-second example. We call this technique \textit{Shuffled mixup} because it shuffles the order of many small chunks in addition to mixing examples together. This expands the \textit{mixup} technique to sequence models and creates additional variety and better regularization during training.

With this final configuration, we no longer saw diverging train/evaluation curves. A comparison of these different techniques can be seen in Table~\ref{data-ablation}.

\begin{table}[ht]
\centering
\begin{tabular}{lcccc}
\hline
Model &  Valid & Test & IDMT & ENST \\
\hline
\textit{Shuffled mixup} & \textbf{88.71}  & \textbf{83.40} & \textbf{85.72} & \textbf{76.89} \\
\textit{mixup} & 79.48 & 69.11 & 47.44 & 62.27 \\
Unmodified & 74.66 & 63.07 & 52.74 & 67.35 \\
\hline
\end{tabular}
\caption{Data augmentation and regularization ablation study. Results are F-measure scores calculated on E-GMD Validation, E-GMD Test, IDMT, and ENST. \textit{Shuffled mixup} is the technique used when training our final OaF-Drums model. Training setup for the other methods is otherwise the same except that training was stopped after approximately 250k steps.}
\label{data-ablation}
\end{table}

After resolving the issue of overfitting to sequences, we also performed a coarse hyperparameter search and discovered that using a smaller convolutional stack prevented the model from overfitting to the particular characteristics of the drum sets in our training dataset. We reduced the number of filters in the convolutional layers from 32/32/64 to 16/16/32 and decreased the units in the fully connected layer from 512 to 256. 

Our final model was trained with a batch size of 128 for 569,400 steps on 16 TPUv3 cores, which took about 3 days. We used the Adam optimizer with an initial learning rate of $1\mathrm{e}{-4}$ and an exponential learning rate decay, reducing by a factor of .98 every 10,000 steps. No early stopping strategy was used other than seeing that the train and evaluation curves had stabilized. We performed a coarse sequential search ($\approx$100 runs) over convolutional architectures, layer sizes, and input resolutions to arrive at the configuration used in the paper.

Code for training and evaluation along with a pre-trained model for inference is available on GitHub: \url{https://goo.gl/magenta/onsets-frames-code}.

\begin{table*}[ht]
\centering
\begin{tabular}{lc | cccc | c}
\multicolumn{1}{c}{} & \multicolumn{1}{c}{} & \multicolumn{4}{c}{} & \multicolumn{1}{c}{Listening} \\
\multicolumn{1}{c}{} & \multicolumn{1}{c}{} & \multicolumn{4}{c}{F-measure} & \multicolumn{1}{c}{Wins} \\
\hline
Model & Training Dataset(s) & IDMT & ENST & E-GMD & E-GMD (vel) & Loop Loft \\
\hline
OaF-Drums & E-GMD & 85.72 & 76.89 & 83.40 & 61.70 & \textbf{919} \\
DT-Ensemble$^*$ & TMIDT(-Bal), MDB, ENST, RBMA  & 91.49 & 82.96  & 64.98 & $\times$ & 677 \\
DT & TMIDT  & $\times$ & 68.00  & $\times$ & $\times$ & $\times$ \\
ADTLib & ENST-3 & 83.12  & $\times$ & $\times$ & $\times$ & 372 \\
\hline
\end{tabular}
\caption{F-measures and listening study results from Section~\ref{listening-test-section}. Note the OaF-Drums model wins the listening study by a significant margin despite achieving comparable classification results to other models. The asterisk on DT-Ensemble$^*$ highlights that the model is actually an ensemble of 5 models trained on 5 different datasets. We use the DT-Ensemble in the listening study as it outperforms the single DT model. OaF-drums is the only model that predicts velocities, so it is the only model to be evaluated on E-GMD velocity labels. Since the various models are trained on different datasets, we compare classifier scores across a range of datasets, and perform the listener studies on the Loop Loft dataset, on which none of the models have been trained.}
\label{big-table}
\end{table*}

\section{Evaluation}
\label{evaluation-section}
Table~\ref{big-table} compares classifier scores for a variety of models and datasets. F-measure (also known as F1 score) is used as the evaluation metric, with a 50ms tolerance window of ground truth annotations for detected onsets as is consistent with the prior studies. We use the \textit{mir\_eval} package for metrics calculation~\cite{raffel2014mir_eval}. 

We compare against the two other models that were also used in the listening study. These models are ADTLib\footnote{\url{https://github.com/CarlSouthall/ADTLib}}
and DrumTranscriptor\footnote{\url{http://ifs.tuwien.ac.at/~vogl/dafx2018/}} (DT), which are from Southall et al.~\shortcite{Southall2017AutomaticDT} and Vogl et al.~\shortcite{multi} respectively. ADTLib is trained on the standard 3 hit ADT task, while DrumTranscriptor is capable of transcribing 18 hits.

The public implementation of DrumTranscriptor is an ensemble 5 models trained on 5 different datasets: TMIDT, TMIDT balanced, ENST, MDB, and RBMA. We refer to this as DrumTranscriptor Ensemble (DT-Ensemble). This contrasts with the single DrumTranscriptor model (DT) in the paper, the best variant of which is trained only on TMIDT. We use DT-Ensemble for our listening study as it outperforms the DT model.

We train OaF-Drums on the E-GMD dataset and evaluate it on IDMT (3-hit standard) and ENST (multi-hit standard) for comparisons to other models.

\subsection{IDMT Evaluation}
IDMT was chosen primarily due to its consistent use in prior studies. It contains only the standard 3 hits (KD, SN, HH). In order to evaluate OaF-Drums in the simpler ADT task, we grouped the 7 possible drum hit predictions into the 3 hits. This grouping is shown in Table~\ref{drum-hierarchy-table}. This is somewhat different than other models we compare against that were trained to predict only those 3 hits and ignore other audio events. We believe this comparison is reasonable because both training/evaluation methods incorporate \textit{a priori} knowledge of what hits need to be predicted. This is yet another example of how different hit mapping strategies makes ADT evaluation difficult. Ultimately, we believe any comparison of models needs to incorporate a perceptual component as we do in the Listening Test in Section~\ref{listening-test-section}.

We evaluated against ADTLib and DT-Ensemble for IDMT. DT-Ensemble uses the same 7 hit grouping that OaF-Drums did. ADTLib only uses the 3 hit grouping and was trained on ENST only considering the standard 3 hits (ENST-3). The IDMT results for ADTLib, OaF-Drums and DT-Ensemble are shown in Table~\ref{big-table}. All models perform rather well, with DT-Ensemble having the best score followed by OaF-Drums.

A full IDMT evaluation against the state of the art models reviewed in Wu et al.~\shortcite{review} is in the Supplement. OaF-Drums has the 3rd best average F-measure of the 11 models. All the other models perform the standard 3 hit classification like ADTLib. The competitive score for OaF-Drums adds confidence that it performs well in the simpler ADT task, especially considering the model has been trained for more complex classification in the number of drum hits and added velocity prediction.        

\subsection{ENST Evaluation}
We evaluate against ENST to compare our model in the multi-hit scenario, beyond the typical 3 hit ADT task.  There are only a few models that attempt to model beyond 3 hits~\cite{Dittmar04furthersteps, multi, choi2019deep, Cartwright2018INCREASINGDT}, and there is no standardization of evaluation for multi-hit models. There are also a very small number of public datasets that have multi-hit annotation, and within those datasets there is inconsistency in number and type of drum hits used.

Of the multi-hit models, Vogl et al.~\shortcite{multi} appear to have the best generalized performance across different datasets, and a public model implementation (DT-Ensemble) was available for additional inference for the listening study. Therefore, we elected to use that work as a proxy for the current state of the art in the multi-hit scenario.

Multi-hit comparison is a non-trivial task since DT-Ensemble is capable of classifying 18 different drum hits, which contrasts to the 25 different drum hits labeled in E-GMD, and the 20 different drum hits labeled in ENST. While there are some consistent mappings between drum hits in each domain, for example, KD, there is a lot of variation and ambiguity in mapping other categories such as cymbals and toms. We elected to evaluate the multi-hit task on a reduction of seven hits shown in Table~\ref{drum-hierarchy-table}. This seven-hit mapping is comparable to the eight-hit model of DT and DT-Ensemble because Clave (the eighth kind of hit) is not used in either our training or evaluation datasets. DT-Ensemble never predicted Clave during evaluation.

\begin{figure}[ht]
\centering
\includegraphics[width=\columnwidth]{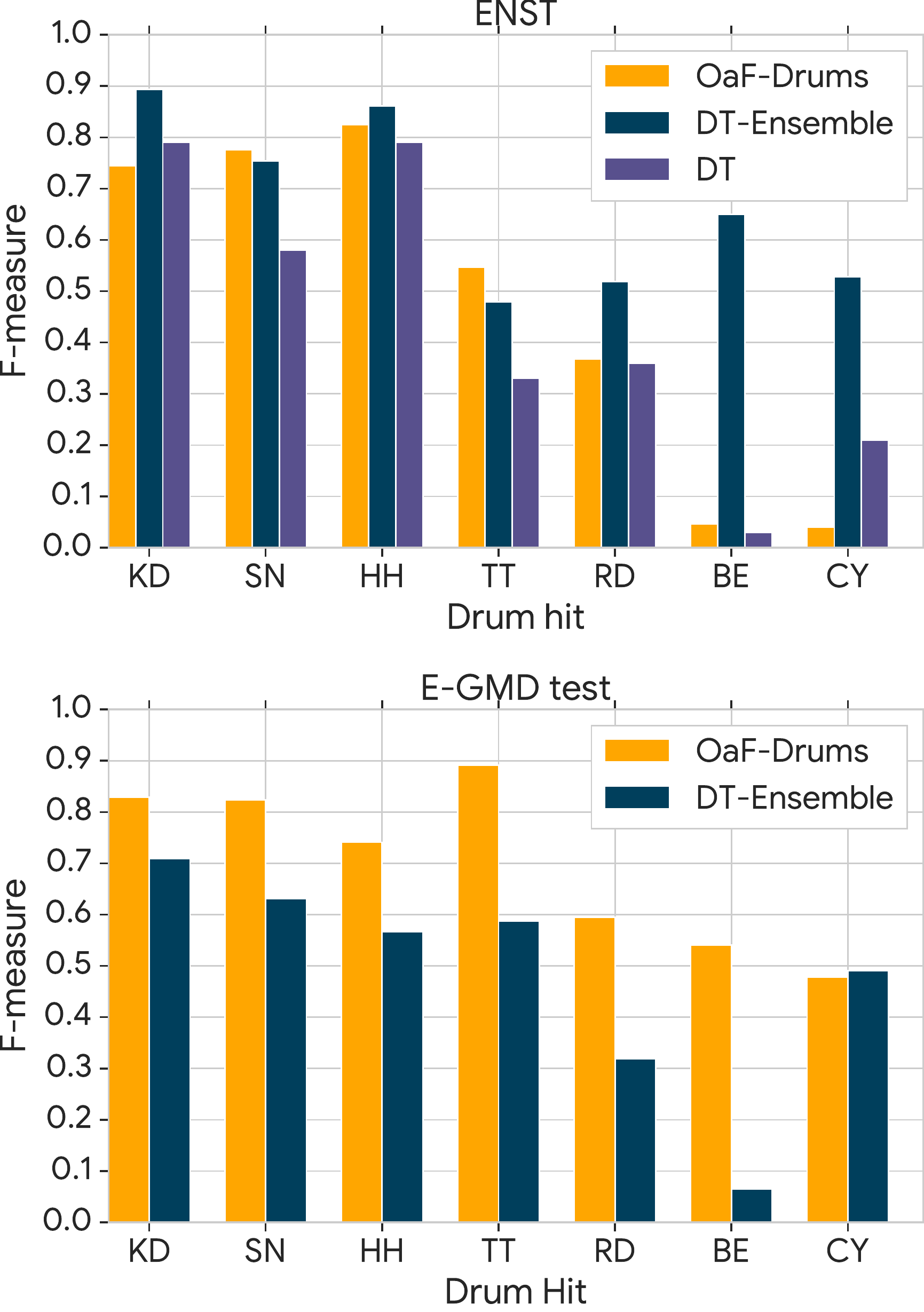}
\caption{The F-measure results per hit on ENST and E-GMD test. The ordering of bars from left to right is OaF-Drums, DT-Ensemble, DT for ENST and OaF-Drums, DT for E-GMD test. DT-Ensemble included ENST in its training set while OaF-Drums and DrumTranscriptor did not. Events such as Bells (BE) are rare and have significant variation between datasets, leading to poor generalization of models not trained on the dataset (OaF-Drums and DT for ENST, and DT-Ensemble for E-GMD).}
\label{enst-per-hit}
\end{figure}

The F-measure results for ENST are shown in Table~\ref{big-table}. OaF-Drums outperforms DT, but both are outperformed by DT-Ensemble, which is expected since DT-Ensemble is trained on ENST. The F-measure results broken down by drum hit are shown in Figure~\ref{enst-per-hit}. 

When broken down by hit, the F-measure results reveal stark contrasts in performance for different hits. Events such as Bells (BE) are rare and have significant variation between datasets, leading to poor generalization of models not trained on the dataset (OaF-Drums and DT for ENST, and DT-Ensemble for E-GMD).

Some attempts have been made to combat this behavior. Applying different weights to onsets in the loss function can help in some cases~\cite{Cartwright2018INCREASINGDT, vogl2017}, but it does not appear effective in the cases of extremely sparse onsets. A more promising approach would be to re-balance the dataset to a more even distribution of onsets, which is explored with the TMIDT dataset in~\cite{multi}. The balanced dataset carried a trade-off in that model however, since per hit F-measures were much more even but overall F-measure notably decreased. 

\subsection{E-GMD Evaluation}
As a final test, we evaluate OaF-Drums and DT-Ensemble against E-GMD. We elect to reduce all drum hit classes down to the same seven classes used in the ENST test as shown in Table~\ref{drum-hierarchy-table}. The results of E-GMD are shown in Table~\ref{big-table}. The F-measure scores for both test and validation are shown in the Supplement.  Not surprisingly, OaF-Drums outperforms DT-Ensemble.  While OaF-Drums did not train on any of the sequences in the E-GMD test subset, the training dataset did have audio from the same drum kits.

We also evaluate OaF-Drums performance using an F-measure score that includes velocity predictions as described in~\cite{hawthorne2018onsets}. We only evaluate OaF-Drums on velocities, as the other models do not predict velocity labels. Results are again shown in Table~\ref{big-table}. Results for both test and validation splits are shown in the Supplement.

Across all datasets, we see that OaF-Drums performs very competitively in an F-measure comparison. This is a good sign of generalization for the model, that it can consistently perform well across datasets not seen during training.

\section{Listening Test}
\label{listening-test-section}
To measure the perceptual quality of our transcription model, we conducted a listening test where raters compared synthesized transcriptions to original recordings. We opted not to use any samples from the standard transcription datasets so that no model would have a particular advantage, and instead used 496 examples drawn from a commercial drum loop set (Loop Loft)\footnote{\url{https://www.thelooploft.com/products/nate-smith-drums-bundle}}. Transcription model outputs were synthesized using FluidSynth\footnote{\url{http://www.fluidsynth.org/}} and the SGMv2.01-Sal-Guit-Bass-V1.3 SoundFont\footnote{\url{https://sites.google.com/site/soundfonts4u/}}. We also decided to focus on comparing models with 7 or fewer output classes because that made it clear how to define a consistent set of General MIDI instruments for synthesis. We mapped all model outputs to the following General MIDI instruments: 36 (Bass Drum 1), 38 (Acoustic Snare), 42 (Closed Hi Hat), 47 (Low-Mid Tom), 49 (Crash Cymbal 1), 51 (Ride Cymbal 1), 53 (Ride Bell).

Synthesizing model output like this has definite limitations. In particular, the drum kit in the SoundFont may sometimes sound very different from the original recording, and velocity changes in the SoundFont typically just scale the volume of the same sample without taking into account the changing physical response of a more or less forceful hit. However, the listening test has the significant advantage of allowing direct comparison of different models in the domain we care about (human perceptual audio similarity) using the same set of sounds.

We compare the outputs of ADTLib, DT-Ensemble, OaF-Drums, and OaF-Drums with output velocities fixed to a constant level. Only OaF-Drums outputs velocity predictions, all others used a fixed velocity of 100.

\begin{figure}[ht]
\centering
\includegraphics[width=\columnwidth]{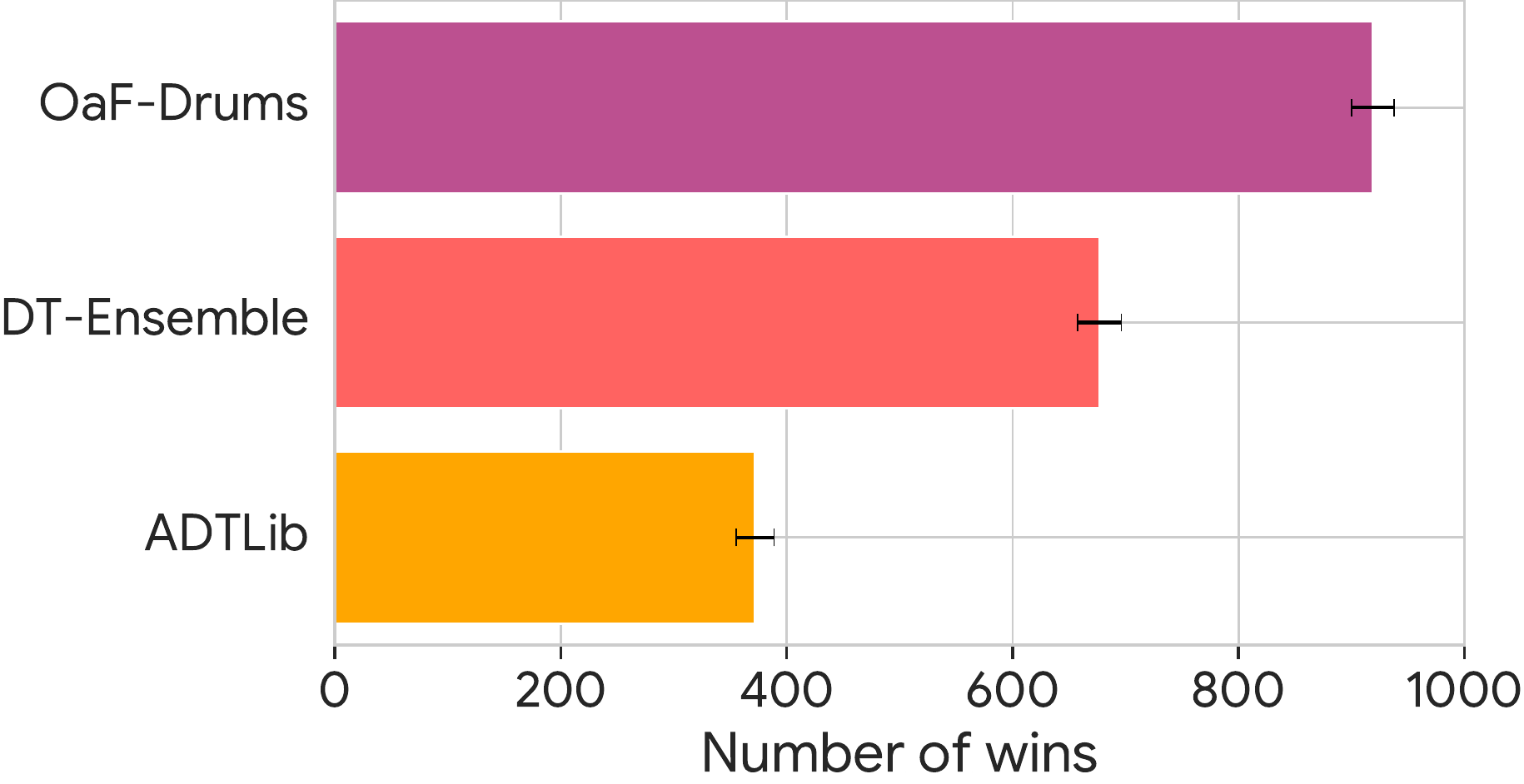}
\caption{Results of our listening tests, showing the number of times each model won in a pairwise comparison. Black error bars indicate estimated standard deviation of means.}
\label{listening-test}
\end{figure}

For each of the 496 examples, we selected a random 10-second clip (or the entire example if it was less than 10 seconds) and the associated synthesized outputs from each of the models. We then generated questions for each of the 6 possible pairwise comparisons between the models, resulting in a total of 2,976 questions. For each question, we asked raters which output best captured the content of the original clip and asked them to rate their choice on a 5-point Likert scale. Figure~\ref{listening-test} shows the number of comparisons in which each source was preferred, with the OaF-Drums model having the overall highest number of wins.

Table~\ref{listening-test-velocity} shows the results of comparing our model with and without velocity predictions and clearly demonstrates the perceptual importance of velocity.

\begin{table}[ht]
\centering
\begin{tabular}{lc}
\hline
Model & Number of wins \\
\hline
OaF-Drums w/ velocity  & 919 \\
OaF-Drums w/o velocity & 456 \\
\hline
\end{tabular}
\caption{Listening test results comparing output of the E-GMD 8 model with velocity predictions and with velocity fixed to a constant level.}
\label{listening-test-velocity}
\end{table}

A Kruskal-Wallis H test of the ratings showed that there is at least one statistically significant difference between the models: $\chi^2(2) = 559.19, p~<~0.001~(7.0846\mathrm{e}{-121})$. A post-hoc analysis using the Wilcoxon signed-rank test with Bonferroni correction showed that there were statistically significant differences between all model pairs with~$p~<~.001/6$.

The online supplement includes examples of listening test comparisons at \url{https://goo.gl/magenta/e-gmd-examples}.

\section{Conclusion and Future Work}
In this work we explored improving perceptual quality in ADT. We introduced the Expanded Groove MIDI Dataset and use the included velocity annotations to train an OaF-Drums model with added velocity predictions. Despite achieving similar results on classification metrics, we showed that multi-hit velocity prediction is well-aligned to the downstream task of generating audio, giving significant improvements in perceptual quality as determined by listening tests.

This work also highlights the value of listening studies in evaluating transcription systems, as an example of classifier outputs as inputs to generative systems. 
Incorporating such studies into the standard suite of classification metrics has the potential to expand the downstream applications of ADT and provide a fair comparison of models between different datasets and architectures. 

Future work could include better representation of more drum hits and combining this model with a pitched automatic music transcription model for full music ensemble transcription.

\bibliography{aaai}

\begin{thebibliography}{29}
\providecommand{\natexlab}[1]{#1}
\providecommand{\url}[1]{\texttt{#1}}
\providecommand{\urlprefix}{URL }
\expandafter\ifx\csname urlstyle\endcsname\relax
  \providecommand{\doi}[1]{doi:\discretionary{}{}{}#1}\else
  \providecommand{\doi}{doi:\discretionary{}{}{}\begingroup
  \urlstyle{rm}\Url}\fi

\bibitem[{Cartwright(2018)}]{Cartwright2018INCREASINGDT}
Cartwright, M. 2018.
\newblock Increasing Drum Transcription Vocabulary Using Data Synthesis.
\newblock In \emph{Proceedings of the International Conference on Digital Audio
  Effects (DAFx)}.

\bibitem[{Choi and Cho(2019)}]{choi2019deep}
Choi, K.; and Cho, K. 2019.
\newblock Deep Unsupervised Drum Transcription.
\newblock In \emph{Proceedings of the International Society for Music
  Information Retrieval Conference (ISMIR), Delft, Netherland}.

\bibitem[{Dittmar and G{\"a}rtner(2014{\natexlab{a}})}]{dittmar2014real}
Dittmar, C.; and G{\"a}rtner, D. 2014{\natexlab{a}}.
\newblock Real-Time Transcription and Separation of Drum Recordings Based on
  NMF Decomposition.
\newblock In \emph{DAFx}, 187--194.

\bibitem[{Dittmar and G{\"a}rtner(2014{\natexlab{b}})}]{Dittmar2014RealTimeTA}
Dittmar, C.; and G{\"a}rtner, D. 2014{\natexlab{b}}.
\newblock Real-Time Transcription and Separation of Drum Recordings Based on
  NMF Decomposition.
\newblock In \emph{DAFx}.

\bibitem[{Dittmar and Uhle(2004)}]{Dittmar04furthersteps}
Dittmar, C.; and Uhle, C. 2004.
\newblock Further steps towards drum transcription of polyphonic music.
\newblock In \emph{in Proc. 11th AES Conv}.

\bibitem[{Gillet and Richard(2006)}]{enst}
Gillet, O.; and Richard, G. 2006.
\newblock ENST-Drums: an extensive audio-visual database for drum signals
  processing.
\newblock In \emph{Proc. Intl. Society for Music Information Retrieval Conf.},
  156--159. ISMIR.

\bibitem[{{Gillick} et~al.(2019){Gillick}, {Roberts}, {Engel}, {Eck}, and
  {Bamman}}]{2019arXiv190506118G}
{Gillick}, J.; {Roberts}, A.; {Engel}, J.; {Eck}, D.; and {Bamman}, D. 2019.
\newblock {Learning to Groove with Inverse Sequence Transformations}.
\newblock \emph{arXiv e-prints} arXiv:1905.06118.

\bibitem[{Hawthorne et~al.(2018)Hawthorne, Elsen, Song, Roberts, Simon, Raffel,
  Engel, Oore, and Eck}]{hawthorne2018onsets}
Hawthorne, C.; Elsen, E.; Song, J.; Roberts, A.; Simon, I.; Raffel, C.; Engel,
  J.; Oore, S.; and Eck, D. 2018.
\newblock Onsets and frames: Dual-objective piano transcription.
\newblock In \emph{Proceedings of the 19th International Society for Music
  Information Retrieval Conference}.

\bibitem[{Lindsay-Smith, McDonald, and Sandler(2012)}]{lindsay2012drumkit}
Lindsay-Smith, H.; McDonald, S.; and Sandler, M. 2012.
\newblock Drumkit transcription via convolutive NMF.
\newblock In \emph{International Conference on Digital Audio Effects (DAFx),
  York, UK}.

\bibitem[{Merity, Keskar, and Socher(2017)}]{merity2017regularizing}
Merity, S.; Keskar, N.~S.; and Socher, R. 2017.
\newblock Regularizing and optimizing LSTM language models.
\newblock \emph{arXiv preprint arXiv:1708.02182} .

\bibitem[{Miron, Davies, and Gouyon(2013)}]{Miron2013AnOD}
Miron, M.; Davies, M. E.~P.; and Gouyon, F. 2013.
\newblock An open-source drum transcription system for Pure Data and Max MSP.
\newblock \emph{2013 IEEE International Conference on Acoustics, Speech and
  Signal Processing} 221--225.

\bibitem[{{Miron}, {Davies}, and {Gouyon}(2013)}]{Miron2013}
{Miron}, M.; {Davies}, M. E.~P.; and {Gouyon}, F. 2013.
\newblock An open-source drum transcription system for Pure Data and Max MSP.
\newblock In \emph{2013 IEEE International Conference on Acoustics, Speech and
  Signal Processing}, 221--225.
\newblock ISSN 2379-190X.
\newblock \doi{10.1109/ICASSP.2013.6637641}.

\bibitem[{Raffel et~al.(2014)Raffel, McFee, Humphrey, Salamon, Nieto, Liang,
  Ellis, and Raffel}]{raffel2014mir_eval}
Raffel, C.; McFee, B.; Humphrey, E.~J.; Salamon, J.; Nieto, O.; Liang, D.;
  Ellis, D.~P.; and Raffel, C.~C. 2014.
\newblock mir\_eval: A transparent implementation of common MIR metrics.
\newblock In \emph{In Proceedings of the 15th International Society for Music
  Information Retrieval Conference, ISMIR}. Citeseer.

\bibitem[{Southall, Stables, and Hockman(2017)}]{Southall2017AutomaticDT}
Southall, C.; Stables, R.; and Hockman, J. 2017.
\newblock Automatic Drum Transcription for Polyphonic Recordings Using Soft
  Attention Mechanisms and Convolutional Neural Networks.
\newblock In \emph{ISMIR}.

\bibitem[{Southall, Stables, and
  Hockman(2018{\natexlab{a}})}]{Carl_Southall_2018}
Southall, C.; Stables, R.; and Hockman, J. 2018{\natexlab{a}}.
\newblock Improving Peak-picking Using Multiple Time-step Loss Functions.
\newblock In \emph{19th International Society for Music Information Retrieval
  Conference}. ISMIR.

\bibitem[{Southall, Stables, and
  Hockman(2018{\natexlab{b}})}]{southall2018player}
Southall, C.; Stables, R.; and Hockman, J. 2018{\natexlab{b}}.
\newblock Player Vs Transcriber: A Game Approach To Data Manipulation For
  Automatic Drum Transcription.
\newblock In \emph{ISMIR}, 58--65.

\bibitem[{Southall et~al.(2017)Southall, Wu, Lerch, and
  Hockman}]{southall2017mdb}
Southall, C.; Wu, C.-W.; Lerch, A.; and Hockman, J. 2017.
\newblock MDB Drums: An annotated subset of MedleyDB for automatic drum
  transcription.
\newblock In \emph{Extended abstracts for the Late-Breaking Demo Session of the
  18th International Society for Music Information Retrieval Conference}.

\bibitem[{Stables, Hockman, and Southall(2016)}]{stables2016automatic}
Stables, R.; Hockman, J.; and Southall, C. 2016.
\newblock Automatic Drum Transcription using Bi-directional Recurrent Neural
  Networks.
\newblock In \emph{17th International Society for Music Information Retrieval
  Conference}.

\bibitem[{Ueda et~al.(2019)Ueda, Shibata, Wada, Nishikimi, Nakamura, and
  Yoshii}]{ueda2019bayesian}
Ueda, S.; Shibata, K.; Wada, Y.; Nishikimi, R.; Nakamura, E.; and Yoshii, K.
  2019.
\newblock Bayesian Drum Transcription Based on Nonnegative Matrix Factor
  Decomposition with a Deep Score Prior.
\newblock In \emph{ICASSP 2019-2019 IEEE International Conference on Acoustics,
  Speech and Signal Processing (ICASSP)}, 456--460. IEEE.

\bibitem[{Vogl, Dorfer, and Knees(2016)}]{vogl2016recurrent}
Vogl, R.; Dorfer, M.; and Knees, P. 2016.
\newblock Recurrent Neural Networks for Drum Transcription.
\newblock In \emph{ISMIR}, 730--736.

\bibitem[{Vogl, Dorfer, and Knees(2017)}]{vogl2017drum}
Vogl, R.; Dorfer, M.; and Knees, P. 2017.
\newblock Drum transcription from polyphonic music with recurrent neural
  networks.
\newblock In \emph{2017 IEEE International Conference on Acoustics, Speech and
  Signal Processing (ICASSP)}, 201--205. IEEE.

\bibitem[{{Vogl}, {Dorfer}, and {Knees}(2017)}]{vogl2017}
{Vogl}, R.; {Dorfer}, M.; and {Knees}, P. 2017.
\newblock Drum transcription from polyphonic music with recurrent neural
  networks.
\newblock In \emph{2017 IEEE International Conference on Acoustics, Speech and
  Signal Processing (ICASSP)}, 201--205.
\newblock ISSN 2379-190X.
\newblock \doi{10.1109/ICASSP.2017.7952146}.

\bibitem[{Vogl et~al.(2017)Vogl, Dorfer, Widmer, and Knees}]{Vogl2017DrumTV}
Vogl, R.; Dorfer, M.; Widmer, G.; and Knees, P. 2017.
\newblock Drum Transcription via Joint Beat and Drum Modeling Using
  Convolutional Recurrent Neural Networks.
\newblock In \emph{ISMIR}.

\bibitem[{Vogl, Widmer, and Knees(2018)}]{multi}
Vogl, R.; Widmer, G.; and Knees, P. 2018.
\newblock Towards Multi-Instrument Drum Transcription.
\newblock In \emph{the 21st International Conference on Digital Audio Effects}.
  DAFx.

\bibitem[{Von~Neumann, Morgenstern, and Kuhn(2007)}]{von2007theory}
Von~Neumann, J.; Morgenstern, O.; and Kuhn, H.~W. 2007.
\newblock \emph{Theory of games and economic behavior (commemorative edition)}.
\newblock Princeton university press.

\bibitem[{Wu et~al.(2018)Wu, Dittmar, Southall, Vogl, Widmer, Hockman, Muller,
  and Lerch}]{review}
Wu, C.-W.; Dittmar, C.; Southall, C.; Vogl, R.; Widmer, G.; Hockman, J.;
  Muller, M.; and Lerch, A. 2018.
\newblock A Review of Automatic Drum Transcription.
\newblock \emph{IEEE/ACM Transactions on Audio, Speech, and Language
  Processing} 26(9): 1457–1483.
\newblock ISSN 2329-9304.
\newblock \doi{10.1109/taslp.2018.2830113}.
\newblock \urlprefix\url{http://dx.doi.org/10.1109/TASLP.2018.2830113}.

\bibitem[{Wu and Lerch(2015)}]{wu2015drum}
Wu, C.-W.; and Lerch, A. 2015.
\newblock Drum Transcription Using Partially Fixed Non-Negative Matrix
  Factorization with Template Adaptation.
\newblock In \emph{ISMIR}, 257--263.

\bibitem[{Wu and Lerch(2018)}]{chih_wei_wu_2018_1492447}
Wu, C.-W.; and Lerch, A. 2018.
\newblock {From Labeled to Unlabeled Data – On the Data Challenge in
  Automatic Drum Transcription}.
\newblock In \emph{{Proceedings of the 19th International Society for Music
  Information Retrieval Conference}}, 445--452. Paris, France: ISMIR.
\newblock \doi{10.5281/zenodo.1492447}.
\newblock \urlprefix\url{https://doi.org/10.5281/zenodo.1492447}.

\bibitem[{Zhang et~al.(2017)Zhang, Cisse, Dauphin, and
  Lopez-Paz}]{zhang2017mixup}
Zhang, H.; Cisse, M.; Dauphin, Y.~N.; and Lopez-Paz, D. 2017.
\newblock mixup: Beyond empirical risk minimization.
\newblock \emph{arXiv preprint arXiv:1710.09412} .

\end{thebibliography}

\newpage
\onecolumn
\section*{Supplement}

\begin{table}[h]
\centering
\begin{tabular}{lcccr}
\hline
Model & Avg & KD & SN & HH \\
\hline
NMFD    & 90.25 & 95.87 & 83.41 & 91.47 \\
SANMF   & 86.53 & 96.40 & 71.70 & 91.50 \\
OaF-Drums & 85.72 & 90.21 & 78.82 & 84.87 \\
GRUts   & 85.14 & 92.49 & 70.30 & 92.64 \\
tanhB   & 84.69 & 96.69 & 69.38 & 87.99 \\
lstmpB  & 83.12 & 96.16 & 70.24 & 82.95 \\ 
PFNMF   & 83.02 & 94.78 & 76.13 & 78.15 \\
RNN     & 80.92 & 88.82 & 61.14 & 92.78 \\
ReLUts  & 80.54 & 91.47 & 58.97 & 91.29 \\
AM1     & 79.69 & 95.91 & 81.16 & 62.00 \\
AM2     & 79.48 & 92.45 & 78.35 & 67.63 \\
\hline
\end{tabular}
\caption{F-measure performance against IDMT, showing the average, and per-instrument performance. The table is sorted in order of best average F-measure performance. Scores for models other than OaF-Drums are from the ``eval cross" experiment described in Wu et al.~\shortcite{review}.}
\label{idmt-table}
\end{table}

\begin{table}[h]
\centering
\begin{tabular}{lcccr}
\hline
Model & Validation & Test \\
\hline 
OaF-Drums & 88.71  & 83.40 \\
DT-Ensemble & 64.07  & 63.98 \\
\hline
\end{tabular}
\caption{F-measure performance against E-GMD validation and test. }
\label{egmd-table}
\end{table}

\begin{table}[h]
\centering
\begin{tabular}{lcccr}
\hline
Model & Validation (Velocity) & Test (Velocity) \\
\hline 
OaF-Drums & 64.97  & 61.70  \\
\hline
\end{tabular}
\caption{F-measure performance including velocity prediction accuracy against E-GMD validation and test. Only OaF-Drums scores are calculated because it is the only model that predicts velocity.}
\label{egmd-velocity-table}
\end{table}

\begin{figure}[h]
\centering
\includegraphics[width=.6\columnwidth]{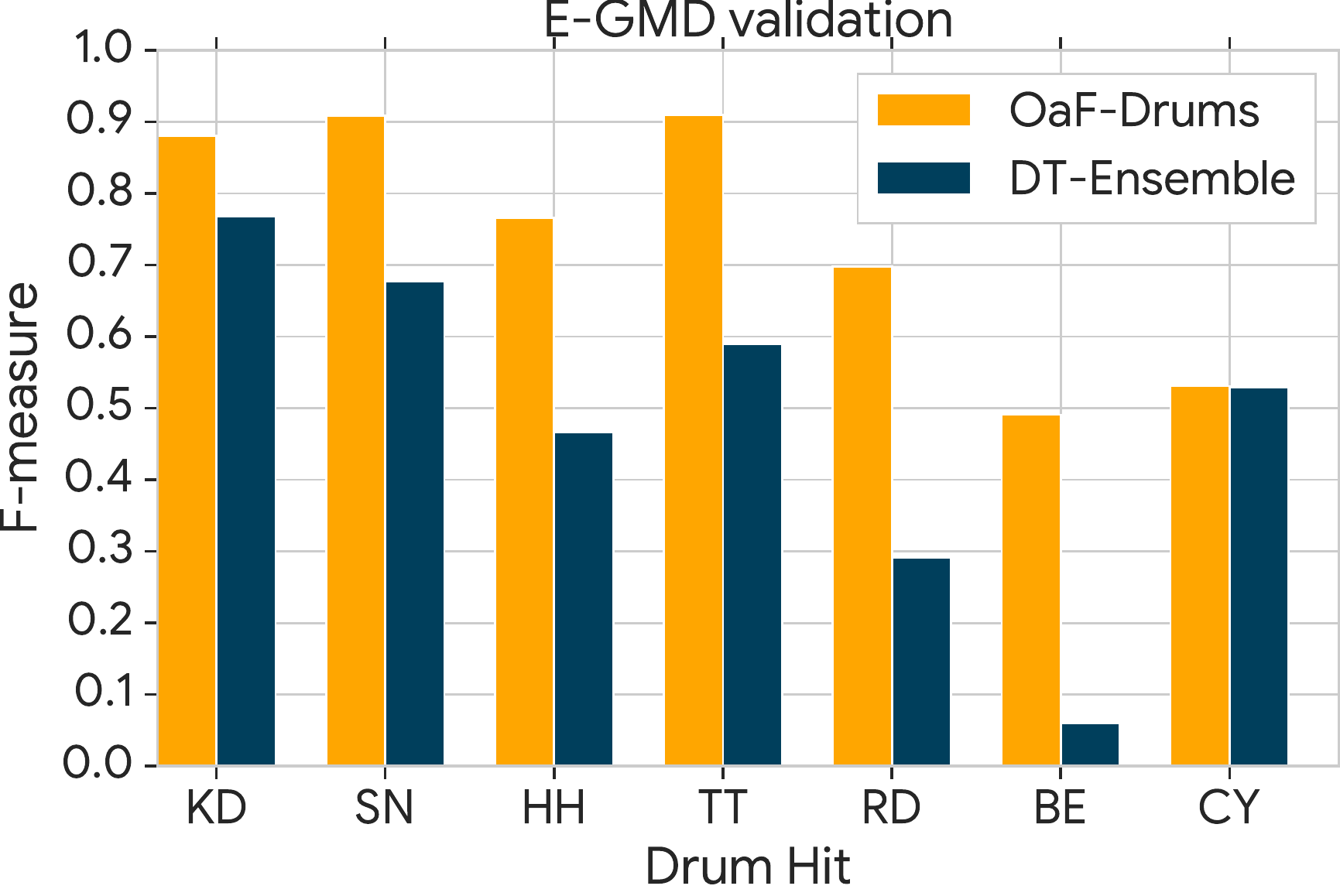}
\caption{The F-measure results per hit on E-GMD validation splits. The ordering of bars from left is OaF-Drums, DT-Ensemble.}
\label{egmd-per-hit}
\end{figure}

\begin{table}[h]
\centering
\begin{tabular}{lccc}
\hline
Layer & Size & Filters & Stride \\
\hline 
Log Mel Spectrogram & 250 bins & & \\
Conv & 16 & 3x3 & 1x1 \\
BatchNorm & & & \\
Conv & 16 & 3x3 & 1x1 \\
BatchNorm &  & & \\
MaxPool & & 1x2 & 1x2 \\
Dropout & & Keep 25\% & \\
Conv & 32 & 3x3 & 1x1 \\
BatchNorm & & & \\
MaxPool & & 1x2 & 1x2 \\
Dropout & & Keep 25\% & \\
Dense & 256 & &  \\
Dropout & & Keep 50\% & \\
Bidirectional LSTM & 64 \\
LSTM Dropout & & Keep 50\% \\
Dense & 88 \\
Sigmoid Cross Entropy & \\
\hline
\end{tabular}
\caption{Onset prediction architecture}
\label{onset-prediction-architecture}
\end{table}

\begin{table}[h]
\centering
\begin{tabular}{lccc}
\hline
Layer & Size & Filters & Stride \\
\hline 
Log Mel Spectrogram & 250 bins & & \\
Conv & 16 & 3x3 & 1x1 \\
BatchNorm & & & \\
Conv & 16 & 3x3 & 1x1 \\
BatchNorm &  & & \\
MaxPool & & 1x2 & 1x2 \\
Dropout & & Keep 25\% & \\
Conv & 32 & 3x3 & 1x1 \\
BatchNorm & & & \\
MaxPool & & 1x2 & 1x2 \\
Dropout & & Keep 25\% & \\
Dense & 256 & &  \\
Dropout & & Keep 50\% & \\
Dense & 88 \\
Mean Squared Error & \\
\hline
\end{tabular}
\caption{Velocity prediction architecture}
\label{velocity-prediction-architecture}
\end{table}

\end{document}